\documentclass[12pt,a4paper]{article}
\usepackage{graphicx,amssymb,amsmath}
\usepackage{setspace} 
\usepackage[linkcolor={blue},citecolor={blue},colorlinks=true,linktocpage,urlcolor={blue}]
{hyperref}
\usepackage{cite}
\usepackage[margin=2cm]{geometry}
\usepackage[dvipsnames]{xcolor}
\definecolor{darkblue}{rgb}{0,0,0.54}
\usepackage{multirow} 
\usepackage{appendix}
\usepackage{pdfpages}
\usepackage{tabu} 
\usepackage{adjustbox} 
\usepackage{braket}
\usepackage[skip=2pt]{caption}
\usepackage{color,soul}
\usepackage[colorinlistoftodos]{todonotes}
\usepackage[normalem]{ulem}
\usepackage{authblk}
\usepackage{bm} 
\usepackage[symbol]{footmisc}

    \makeatletter
\def\@fnsymbol#1{\ensuremath{\ifcase#1\or \dagger\or \ddagger\or
   \mathsection\or \mathparagraph\or \|\or **\or \dagger\dagger
   \or \ddagger\ddagger \else\@ctrerr\fi}}
    \makeatother
\graphicspath{{Figures/}} 

\title{Revealing the Berry phase under the tunneling barrier}
\global\long\def\at#1{\left.#1\right|}%

	\author[1]{Lior Faeyrman}
 	\author[2]{Eduardo B. Molinero}
	\author[1]{Roni Weiss}
	\author[3]{Vladimir Narovlansky}
 		\author[1]{Omer Kneller}
	\author[1]{Talya Arusi-Parpar}
	\author[1]{Barry D. Bruner}
	\author[4]{Binghai Yan}
	\author[5,7,8]{Misha Ivanov}
 	\author[5,6]{Olga Smirnova}
 		\author[2]{\'Alvaro Jim\'enez-Gal\'an}
    	\author[9]{Riccardo Piccoli}
 	\author[5,2]{Rui E.F. Silva}
	\author[1]{Nirit Dudovich $^{*}$}
	\author[3]{Ayelet J. Uzan-Narovlansky $^{*}$}

	\affil[1]{\footnotesize Department of Complex Systems, Weizmann Institute of Science, 76100, Rehovot, Israel}
 			\affil[2]{\footnotesize Instituto de Ciencia de Materiales de Madrid (ICMM), Consejo Superior de Investigaciones Científicas (CSIC), Madrid, Spain}
	\affil[3]{\footnotesize Department of Physics, Princeton University, Princeton, NJ 08544, USA}
		\affil[4]{\footnotesize Department of Condensed Matter, Weizmann Institute of Science, 76100, Rehovot, Israel}
	\affil[5]{\footnotesize Max-Born-Institut, Max-Born Strasse 2A, D-12489 Berlin, Germany}
		\affil[6]{\footnotesize Technische Universit\"at Berlin, Ernst-Ruska-Geb\"aude, Hardenbergstr. 36A, D-10623 Berlin, Germany}

\affil[7]{\footnotesize Blackett Laboratory, Imperial College London, South Kensington Campus, SW7 2AZ London, United Kingdom}
	\affil[8]{\footnotesize Department of Physics, Humboldt University, Newtonstrasse 15, 12489 Berlin, Germany}
 	\affil[9]{\footnotesize Department of Molecular Sciences and Nanosystems, Ca' Foscari University of Venice, via Torino 155, 30172 Venice, Italy}

 \footnotetext{Corresponding authors: nirit.dudovich@weizmann.ac.il, auzan@princeton.edu}

\begin{document}
\maketitle

\begin{abstract}

In quantum mechanics, a quantum wavepacket may acquire a geometrical phase as it evolves along a cyclic trajectory in parameter space \cite{berry_quantal_1984}. In condensed matter systems, the Berry phase plays a crucial role in fundamental phenomena such as the Hall effect, orbital magnetism, and polarization \cite{xiao2010berry}. Resolving the quantum nature of these processes commonly requires sensitive quantum techniques, as tunneling, being the dominant mechanism in STM microscopy and tunneling transport devices. In this study, we integrate these two phenomena – geometrical phases and tunneling – and observe a complex-valued Berry phase via strong field light matter interactions in condensed matter systems\cite{ghimire2019high}. By manipulating the tunneling barrier\cite{kneller2022look}, with attoseconds precision, we measure the imaginary Berry phase accumulated as the electron tunnels during a fraction of the optical cycle. Our work opens new theoretical and experimental directions in geometrical phases physics and their realization in condensed matter systems, expanding solid state strong field light metrology to study topological quantum phenomena.

\end{abstract}

\newpage
\section{MAIN}

When a quantum system evolves in a cyclic path in parameter space, it accumulates a geometrical phase attributed to the parameter space
topology. This phase, with its formulations of the Aharonov-Bohm phase
 \cite{aharonov_significance_1959} and the Berry phase \cite{berry_quantal_1984}, has been
at the heart of topological phenomena in a broad range of fields - from condensed matter physics
\cite{xiao_berry_2010,dutreix_measuring_2019}, fluid mechanics \cite{berry_wavefront_1980}, optics \cite{pancharatnam_generalized_1956,simon_evolving_1988}, 
to particle physics \cite{wilson1974confinement}. The role of the Berry phase is essential in many solid state systems, as it appears in the electronic Bloch states \cite{cohen2019geometric,xiao2010berry} and leads
to a wide range of observations such as the quantum Hall effect, macroscopic
polarization \cite{king-smith_theory_1993}, orbital magnetism
and the topological classification of materials \cite{hasan_colloquium_2010}.

The geometric phase has been commonly studied in closed quantum systems, resulting in real-valued phases. However, an interesting and unique scenario may arise in open systems, where the wavefunction is coupled to the environment. This picture can be formally described as an evolution of the electronic wavefunction following a path in a complex parameter space. Here, the geometric phase itself becomes complex, acquiring an imaginary component which represents an amplification or attenuation of the wavefunction's amplitude. The notion of a complex geometric phase for open systems was previously introduced in non hermitian systems \cite{cohen_geometric_2019} as spin systems with geometric dephasing \cite{whitney_geometric_2005,leek2007observation}. This concept became extremely important in the fast evolving field of quantum information, reflecting the topological dephasing of the information \cite{carollo2003geometric,duan2001geometric,jones2000geometric}. Despite the important role that the complex geometrical phase may play in a wide range of systems, previous studies have mainly focused on its real component.

In this paper, we introduce and experimentally observe, for the first time, the notion
of a complex geometric phase arising in condensed matter systems, by linking this phase with an additional quantum process -- tunneling. Integrating these two quantum phenomena introduces a unique physical system to observe the complex Berry phase. We can describe the origin of this complex phase according to the following schematic picture. Consider a quantum wavefunction, propagating in a closed loop in parameter space, accumulating a real Berry phase (figure 1a). Introducing a small tunneling barrier, raises the following question: will the the geometric properties of the system leave their fingerprint on the wavefunction as it propagates under the barrier? The evolution of the electronic wavefunction under the tunneling barrier can be described using complex time and complex momentum, which are the hallmark of tunneling mechanism \cite{keldysh1965ionization,ivanov_ionization_2014,pedatzur2015attosecond,eckle2008attosecond}. Such evolution along these complex parameters leads to the accumulation of a complex geometrical phase. The imaginary component of this phase represents a dissipation of the wavepacket amplitude, dictated by the path it took in parameter space under the barrier.
Interestingly, previous studies have shown that the imaginary Berry phase is gauge-independent, removing the requirement of a closed loop \cite{whitney_geometric_2005} (see SI). Moreover, we find that the imaginary component has a significant contribution to the well known and measurable quantities in quantum geometry, such as the Berry curvature and the quantum metric \cite{gianfrate2020measurement,torma2023essay} (see SI).

The imaginary Berry phase may appear in a wide range of systems. Here we focus on its observation in light-driven condensed matter
systems. In solid state systems, an electron traverses
a closed loop through the Brillouin zone (BZ) in a specific energy
band, leading to the accumulation of a Zak phase \cite{zak_berrys_1989}. The Zak phase admits real values which depend on the topology of the BZ and the band structure. Recently, a generalization of this phase
introduced the interband Berry phase, including both adiabatic and discrete modifications of the wavefunction between the bands \cite{uzan-narovlansky_observation_2024,chacon_circular_2020,yue2020imperfect}. Driven by a low-frequency external field, the electronic
wavepacket undergoes a non-adiabatic interband tunneling transition
followed by intraband propagation and closed by an additional non-adiabatic
transition by photo-recombination\cite{vampa2015linking}. This process, known as high harmonic generation (HHG) leads to the emission of optical radiation with attosecond duration, mapping the accumulated complex phase of the electronic wavefunction into the optical amplitude and phase of the emitted pulses. Solid state HHG has become one of the most promising metrology schemes to resolve a wide variety of condensed matter phenomena \cite{ghimire2011observation,luu2015extreme,schubert2014sub},  providing unique key advantages. The first is associated to its short time scale of the process, which intrinsically occurs within a single optical cycle of the driving laser field.  This allows probing of the system coherence before any scattering or dephasing mechanisms \cite{schultze2014attosecond}. The second is the precise control over the electron trajectories in momentum space, allowing for example valley manipulation with few-femtosecond switching between quantum degrees of freedom \cite{langer2018lightwave,tyulnev2024valleytronics,mitra2024light}.

How do we isolate the role of the imaginary Berry phase and separate it from all additional components of the interaction? Driving the interaction with a two color field enables us to accurately manipulate the origin of imaginary phases in the HHG process -- the tunneling barrier itself (figure 1). We add a weak second harmonic field (SH), polarized parallel to the fundamental field, to enhance or to suppress the tunneling barrier \cite{kneller2022look,pedatzur2015attosecond}. Modulating the tunneling barrier reveals the role of the imaginary Berry phase and allows the reconstruction of its geometrical properties.
In the following we take a deeper look into the formation of the imaginary Berry phase in light driven crystals. In the absence of the perturbation (the SH field), the emitted field with its phase accumulated by the electron-hole wavepacket is
given by:
\begin{equation}
E_{HHG}(t)\propto e^{iS\left[\boldsymbol{k},t_0,t\right]}=e^{i\int_{t_0}^{t}[\varepsilon_{g}\left(\boldsymbol{k}\left(t'\right)\right)+\boldsymbol{F}\left(t'\right)\cdot\left(\boldsymbol{\mathcal{A}}_{g}\left(\boldsymbol{k}\left(t'\right)\right)+\nabla_{\boldsymbol{k}}\phi_{d}\left(\boldsymbol{k}\left(t'\right)\right)\right)]dt'}\label{eq:XUV phase}
\end{equation}
with $S\left[\boldsymbol{k},t_0,t\right]$ being the semi-classical
action, $\boldsymbol{k}$ recombination quasi-momentum, $t_0$ and $t$ are the tunneling and recombination times, $\varepsilon_{g}=\varepsilon_{c}-\varepsilon_{v}$
 the conduction-valence band gap, $\boldsymbol{F}=-\partial_{t}\boldsymbol{A}$
the external electric field, $\boldsymbol{A}$ the vector potential, $\boldsymbol{\mathcal{A}}_{g}=\boldsymbol{\mathcal{A}}_{c}-\boldsymbol{\mathcal{A}}_{v}$
the electron-hole relative Berry connection $\boldsymbol{\mathcal{A}}_{n}\left(\boldsymbol{k}\right)=i\left\langle \at{u_{n,\boldsymbol{k}}}\nabla_{\boldsymbol{k}}u_{n,\boldsymbol{k}}\right\rangle $,
$\phi_{d}$ the transition dipole phase $\phi_{d}\left(\boldsymbol{k}\right)=arg\left(i\left\langle \at{u_{c,\boldsymbol{k}}}\nabla_{\boldsymbol{k}}u_{v,\boldsymbol{k}}\right\rangle \right)$,
$\boldsymbol{k}\left(\tau\right)=\boldsymbol{k}-\boldsymbol{A}\left(t\right)+\boldsymbol{A}\left(t'\right)$
being the crystal quasi-momentum. The phase in equation (\ref{eq:XUV phase})
was previously shown to be gauge invariant \cite{chacon_circular_2020,uzan-narovlansky_observation_2024,yue2020imperfect}, and thus observable. While the electron-hole wavepacket propagates inside the classically
forbidden region, it experiences complex time and accumulates complex
momentum, leading the geometric phase to take complex values. A further look into the action
in (\ref{eq:XUV phase}) reveals that the geometric contribution, given by $\boldsymbol{F}\cdot\boldsymbol{\mathcal{A}}_g$,
acts as an effective modulation of the band gap $\varepsilon_{g}^{eff}=\varepsilon_{g}+\boldsymbol{F}\cdot\boldsymbol{\mathcal{A}}_g$.
Such modification leads to a direct enhancement or suppression of the tunneling probability amplitude according to: $Im\left\{ \int_{t_0^r+it_0^i}^{t_0^r}\boldsymbol{F}\left(t'\right)\cdot\boldsymbol{\mathcal{A}}_{g}\left(\boldsymbol{k}\left(t'\right)\right)dt'\right\} $ with $t_0^r$ and $t_0^i$ being the real and imaginary ionization times\cite{kneller_look_2022}. 
Here, the accumulated
Berry phase is the integration of the Berry connection along the complex path in parameter space. In order to isolate this phase we have to manipulate the tunneling barrier in a controllable manner. Such control is achieved by adding the SH field, $\boldsymbol{A}=\boldsymbol{A_\omega}+\boldsymbol{A_{2\omega}}$, shaping the subcycle evolution of the field and therefore of the tunnelling barrier itself. Scanning the relative subcycle delay between the two fields, $\boldsymbol{A_{2\omega}}=\boldsymbol{A_{2\omega_0}}\sin[2\omega_{0}(t+\tau)]$ ($\omega_0$ is the fundamental field frequency), while keeping the fundamental field constant, enhances or suppresses the tunneling barrier. This scheme reveals the role of the imaginary Berry phase and allows its reconstruction.

We experimentally demonstrate the observation of the complex Berry phase by producing
HHG in $\alpha$-quartz z-cut sample, driven by $\lambda=1.2\mu m$, $50fs$ laser field having an intensity
of the order of $10^{13}W/cm^{2}$ \cite{garg2018ultimate,hammond2017producing}. $\alpha$-quartz exhibits broken inversion
symmetry that gives rise to a non-vanishing Berry curvature and the
appearance of even harmonics\cite{luu2018measurement}. 
The harmonic spectrum spans up to $30eV$, probing the complex dynamics over a wide energy range. The SH perturbation field is produced
by using a $100\mu m$ type-I phase barium borate (BBO) ($BaB_{2}O_{4}$),
and the delay is scanned with Fused-Silica wedges. A detailed description of the experimental setup is provided at the SI.

Figure 2a presents the HHG  spectrum as a function of the two color delay. The experimental results show two important observations. First, the harmonic signal oscillates at $2\omega_0$, with the two color delay \cite{luu2018observing}. This observation is in contrast with previous studies in gas phase and in inversion symmetric crystals, where $4\omega_0$ oscillations were resolved \cite{dudovich2006measuring,pedatzur2015attosecond,uzan2020attosecond}. Second, in contrast to previous observations of solid state HHG \cite{vampa2015linking}, here even and odd harmonic orders oscillate almost in phase. We can understand these results by performing a perturbative analysis.

The SH field adds a complex perturbation to the action (equation \ref{eq:XUV phase})
$\sigma(\tau)=\sigma^{r}+i\sigma^{i}$ . The real perturbation $\sigma^{r}$, resolves the atto-chirp \cite{dudovich2006measuring,vampa2015linking}, while the imaginary perturbation $\sigma^i$ encodes the tunneling process itself. Due to symmetry, the perturbation inverses its sign between consecutive half cycles of the fundamental laser field, manipulating the relative amplitude and phase of the emitted harmonics. The interference signal is mapped into the signal of odd and even harmonics (see SI):
\begin{equation}
\begin{cases}
I_{odd,N}\left(\sigma\right)\propto e^{-2Im\left(\varepsilon_{g}\right)}\left|e^{-i\gamma^{r}_B-\gamma^{i}_B+i\sigma}+e^{i\gamma^{r}_B+\gamma^{i}_B-i\sigma}\right|^{2}\\
I_{even,N}\left(\sigma\right)\propto e^{-2Im\left(\varepsilon_{g}\right)}\left|e^{-i\gamma^{r}_B-\gamma^{i}_B+i\sigma}-e^{i\gamma^{r}_B+\gamma^{i}_B-i\sigma}\right|^{2}
\end{cases}\label{eq:I_odd I_even}
\end{equation}
where $\varepsilon_{g}$ is the band gap, $\gamma^{r}_B=Re\left\{ \int_{t_0}^{t}\boldsymbol{F}\left(t'\right)\cdot\left(\boldsymbol{\mathcal{A}}_{g}\left(\boldsymbol{k}\left(t'\right)\right)+\nabla_{\boldsymbol{k}}\phi_{d}\left(\boldsymbol{k}\left(t'\right)\right)\right)dt'\right\} $
contains the real interband Berry phase (also known as the Shift Vector
\cite{qian_role_2022}) and $\gamma^{i}_B=Im\left\{ \int_{t_0}^{t}\boldsymbol{F}\left(t'\right)\cdot\boldsymbol{\mathcal{A}}_{g}\left(\boldsymbol{k}\left(t'\right)\right)dt'\right\} $
is the imaginary part of the Berry phase, accumulated under the tunneling
barrier. Expanding equation (\ref{eq:I_odd I_even}) to
first order in $\sigma$ one gets:
\begin{equation}
I_{\pm,N}\left(\sigma\right)\propto     cosh\left(2\gamma^{i}_B\right)\pm cos\left(2\gamma^{r}_B\right)+4sinh\left(2\gamma^{i}\right)\sigma^{i}\pm4sin\left(2\gamma^{r}_B\right)\sigma^{r}+O\left(\sigma^{2}\right)\label{eq:I_HHG first order}
\end{equation}
with '$+$' and $'-'$ for odd and even harmonics, respectively. Importantly, as we scan the two-color delay, $\sigma^{i},\sigma^{r}$
oscillate at $2\omega_0$ frequency, leading to $2\omega_0$ oscillations of the harmonic signal itself. We note that in the case of a zero Berry phase, the first order perturbation becomes zero, leading to second order perturbation oscillating at $4\omega_0$.

Our experimental results reveal that odd and even harmonics oscillate approximately in phase (figure 2a). According to equation (3), such an observation identifies the dominant role of the imaginary perturbation $\sigma^i$ over the real perturbation $\sigma^r$. A careful analysis enables a direct isolation of the real and imaginary components of the perturbation. Extracting the difference between neighboring harmonic orders isolates the real perturbation, $\sigma^r$. Such analysis resolves the well known atto-chirp (see SI in \cite{uzan-narovlansky_observation_2024}), revealing the signature of the interband mechanism \cite{vampa_merge_2017}. Importantly, extracting the sum of neighboring harmonics isolates the imaginary perturbation $\sigma^i$. Figure 2b describes the imaginary component of the perturbation $\sigma^{i}$
as a function of the delay and harmonic number along the $\Gamma-K$ direction. As can be clearly observed, the oscillation phase of the imaginary perturbation is almost flat across the harmonics spectrum. Indeed such a flat spectral response has been identified in gas phase experiments \cite{pedatzur2015attosecond}, resolving the short temporal window of the tunneling mechanism.

The sum of adjacent harmonics, $I_{ sum,N}$, allows the reconstruction of the imaginary Berry phase.
This sum isolates the overall imaginary phase, probed by the SH perturbation -- $I_{ sum,N}\left(\sigma\right)\propto tanh\left(2\gamma^{i}_B\right)\sigma^{i}+O\left(\sigma^{2}\right)$. Extracting the amplitude of the oscillating term, by applying Fourier transformation, retrieves the imaginary Berry phase itself ($\propto tanh\left(2\gamma^{i}_B\right))$. Reconstructing the imaginary Berry phase for different harmonic numbers, maps this property for different recombination momentum along the fundamental polarization axis. Finally, we perform the analysis at various crystal angles, mapping the imaginary Berry phase along the 2D momentum space (figure 3a).

Figure 3c describes $I_{ sum,N}\left(\sigma\right)$ for
harmonics $14,15$ as a function of delay and crystal angle ($\theta$), where
$\Gamma-K$ and $\Gamma-M$ lie on $0^{\circ}$ and $30^{\circ}$
respectively. The signal oscillates in clear $2\omega_0$ frequency,
revealing the non-zero value of the imaginary Berry phase. As we rotate the crystal to $\Gamma-M$ the contrast decreases. Indeed, along this axis the Berry connection, and therefore the Berry phase, vanish. Importantly, at  $\Gamma-K'$ we observe a clear phase shift by $\pi$ with respect to $\Gamma-K$. This shift can be understood by the $C3$ symmetry of the crystal. A $60^{\circ}$ rotation is equal to a $180^{\circ}$ rotation, leading to an inversion of the imaginary Berry phase (seen as inversion of the Berry connection, figure 3b). Consequently, the perturbation phase, which probes such asymmetry, shifts by $\pi$. This observation is consistent with previous THz-resolved HHG measurements in broken inversion crystals\cite{langer2017symmetry}. We have performed extensive simulations (see SI) to confirm the experimental features. In figure 3d, we show the theoretical $I_{sum,N}(\sigma)$ where one can appreciate that the same physics as in the experimental results appears: We first see a clear oscillation of $2\omega_0$ at $0^\circ$, followed by a decrease in the contrast at $30^\circ$ and finally a $\pi$-shifted revival of the $2\omega_0$ oscillations at $60^\circ$.\\

Next, we apply Fourier transformation and isolate the imaginary Berry phase at each crystal orientation. In figure 4a, we present the angular dependence of the imaginary Berry phase, $\propto tanh\left(2\gamma^{i}_B\right) $, for harmonic H13, measured with different laser intensities. First we notice that the crystal symmetry is clearly reflected for all the intensity values -- as we rotate the crystal from $0^\circ$ to $60^\circ$ the imaginary Berry phase shifts by $\pi$. In addition, increasing the field's intensity leads to higher values of the imaginary Berry phase. What is the origin of this observation? As we increase the field we reduce the tunneling barrier and therefore the imaginary component of the time and momentum decreases. However, the Berry phase itself is directly proportional to the vector field which increases with the laser intensity. Clearly, in the balance between the two effects, the second effect plays the dominant role. This picture is well confirmed by our calculations (figure 4c). A detailed discussion of this picture is provided in the SI.

In order to get a deeper insight into the subcycle evolution of the interaction, we resolve the variation of the imaginary Berry phase with the harmonic number. In figure 4b, we plot the reconstructed imaginary Berry phase associated with harmonics H11 to H14, corresponding to the interaction induced from the first conduction band. As in the intensity scan, the crystal symmetry is clearly reflected, as we notice a $\pi$ phase shift around $30^\circ$ for all the harmonic orders. Importantly, we observe that the imaginary Berry phase increases with harmonic number. Higher harmonics are associated with earlier tunneling times, and therefore a higher instantaneous field. Since the Berry phase is proportional to the field's strength, increasing the harmonic order leads to an increase of this phase, revealing its subcycle evolution (see SI).

\subsection*{Summary}

Our study pushes the boundaries of the well-known Berry phase in solids by generalizing this concept into the complex plane, thereby shedding light on geometrical phenomena within the classically forbidden region.  
We demonstrate, both theoretically and experimentally, the ability to isolate the role of the imaginary Berry phase, accumulated as the electron propagates under the tunneling barrier. Applying solid state two colors high harmonic spectroscopy, we manipulate the tunneling barrier with sub-cycle precision, extracting the imaginary Berry phase and identifying its geometrical properties. Resolving the variation of this phase with the crystal orientation captures its symmetrical properties, reflecting the symmetry of the tunneling barrier itself. Following the variation of this phase with the harmonic order, visualizes its evolution within a fraction of the optical cycle. Our study will open new opportunities for exploring novel quantum phenomena in condensed matter systems, such as topology \cite{schmid2021tunable, bai2021high} and magnetism. Furthermore, our proposed scheme will introduce new approaches for controlling the system's geometrical properties by extending them into the complex plane. Our theoretical study of the Berry curvature and the quantum metric will open an opportunity to study quantum geometry in a wide range of open systems. Finally, the capacity to resolve complex geometrical phases extends beyond strong field light-matter interactions into the realms of quantum information and quantum computation  -- exploring dephasing and quantum entanglement phenomena.

\bibliography{main_Ayelet}

\begin{thebibliography}{10}

\bibitem{berry_quantal_1984}
M.~V. Berry.
\newblock Quantal phase factors accompanying adiabatic changes.
\newblock 392(1802):45--57.
\newblock Publisher: The Royal Society.

\bibitem{xiao2010berry}
Di~Xiao, Ming-Che Chang, and Qian Niu.
\newblock Berry phase effects on electronic properties.
\newblock {\em Reviews of modern physics}, 82(3):1959, 2010.

\bibitem{ghimire2019high}
Shambhu Ghimire and David~A Reis.
\newblock High-harmonic generation from solids.
\newblock {\em Nature physics}, 15(1):10--16, 2019.

\bibitem{kneller2022look}
Omer Kneller, Doron Azoury, Yotam Federman, Michael Krueger, Ayelet~J Uzan, Gal
  Orenstein, Barry~D Bruner, Olga Smirnova, Serguei Patchkovskii, Misha Ivanov,
  et~al.
\newblock A look under the tunnelling barrier via attosecond-gated
  interferometry.
\newblock {\em Nature photonics}, 16(4):304--310, 2022.

\bibitem{aharonov_significance_1959}
Y.~Aharonov and D.~Bohm.
\newblock Significance of electromagnetic potentials in the quantum theory.
\newblock 115(3):485--491.
\newblock Publisher: American Physical Society.

\bibitem{xiao_berry_2010}
Di~Xiao, Ming-Che Chang, and Qian Niu.
\newblock Berry phase effects on electronic properties.
\newblock 82(3):1959--2007.
\newblock Publisher: American Physical Society.

\bibitem{dutreix_measuring_2019}
C.~Dutreix, H.~González-Herrero, I.~Brihuega, M.~I. Katsnelson, C.~Chapelier,
  and V.~T. Renard.
\newblock Measuring the berry phase of graphene from wavefront dislocations in
  friedel oscillations.
\newblock 574(7777):219--222.
\newblock Publisher: Nature Publishing Group.

\bibitem{berry_wavefront_1980}
M.~V. Berry, R.~G. Chambers, M.~D. Large, C.~Upstill, and J.~C. Walmsley.
\newblock Wavefront dislocations in the aharonov-bohm effect and its water wave
  analogue.
\newblock 1(3):154.

\bibitem{pancharatnam_generalized_1956}
S.~Pancharatnam.
\newblock Generalized theory of interference, and its applications. part i.
  coherent pencils.
\newblock 44(5):247--262.
\newblock Number: 5 Publisher: Indian Academy of Sciences.

\bibitem{simon_evolving_1988}
R.~Simon, H.~J. Kimble, and E.~C.~G. Sudarshan.
\newblock Evolving geometric phase and its dynamical manifestation as a
  frequency shift: An optical experiment.
\newblock 61(1):19--22.
\newblock Publisher: American Physical Society.

\bibitem{wilson1974confinement}
Kenneth~G Wilson.
\newblock Confinement of quarks.
\newblock {\em Physical review D}, 10(8):2445, 1974.

\bibitem{cohen2019geometric}
Eliahu Cohen, Hugo Larocque, Fr{\'e}d{\'e}ric Bouchard, Farshad Nejadsattari,
  Yuval Gefen, and Ebrahim Karimi.
\newblock Geometric phase from aharonov--bohm to pancharatnam--berry and
  beyond.
\newblock {\em Nature Reviews Physics}, 1(7):437--449, 2019.

\bibitem{king-smith_theory_1993}
R.~D. King-Smith and David Vanderbilt.
\newblock Theory of polarization of crystalline solids.
\newblock 47(3):1651--1654.
\newblock Publisher: American Physical Society.

\bibitem{hasan_colloquium_2010}
M.~Z. Hasan and C.~L. Kane.
\newblock Colloquium: Topological insulators.
\newblock 82(4):3045--3067.
\newblock Publisher: American Physical Society.

\bibitem{cohen_geometric_2019}
Eliahu Cohen, Hugo Larocque, Frédéric Bouchard, Farshad Nejadsattari, Yuval
  Gefen, and Ebrahim Karimi.
\newblock Geometric phase from aharonov–bohm to pancharatnam–berry and
  beyond.
\newblock 1(7):437--449.
\newblock Publisher: Nature Publishing Group.

\bibitem{whitney_geometric_2005}
Robert~S. Whitney, Yuriy Makhlin, Alexander Shnirman, and Yuval Gefen.
\newblock Geometric nature of the environment-induced berry phase and geometric
  dephasing.
\newblock 94(7):070407.
\newblock Publisher: American Physical Society.

\bibitem{leek2007observation}
Peter~J Leek, JM~Fink, Alexandre Blais, R~Bianchetti, M~Goppl, Jay~M Gambetta,
  David~I Schuster, Luigi Frunzio, Robert~J Schoelkopf, and Andreas Wallraff.
\newblock Observation of berry's phase in a solid-state qubit.
\newblock {\em science}, 318(5858):1889--1892, 2007.

\bibitem{carollo2003geometric}
Angelo Carollo, Ivette Fuentes-Guridi, M~Franca Santos, and Vlatko Vedral.
\newblock Geometric phase in open systems.
\newblock {\em Physical review letters}, 90(16):160402, 2003.

\bibitem{duan2001geometric}
L-M Duan, Juan~I Cirac, and Peter Zoller.
\newblock Geometric manipulation of trapped ions for quantum computation.
\newblock {\em Science}, 292(5522):1695--1697, 2001.

\bibitem{jones2000geometric}
Jonathan~A Jones, Vlatko Vedral, Artur Ekert, and Giuseppe Castagnoli.
\newblock Geometric quantum computation using nuclear magnetic resonance.
\newblock {\em Nature}, 403(6772):869--871, 2000.

\bibitem{keldysh1965ionization}
LV~Keldysh et~al.
\newblock Ionization in the field of a strong electromagnetic wave.
\newblock {\em Sov. Phys. JETP}, 20(5):1307--1314, 1965.

\bibitem{ivanov_ionization_2014}
Misha Ivanov.
\newblock Ionization in strong low-frequency fields.
\newblock In {\em Attosecond and {XUV} Physics}, pages 177--200. John Wiley \&
  Sons, Ltd.
\newblock Section: 6 \_eprint:
  https://onlinelibrary.wiley.com/doi/pdf/10.1002/9783527677689.ch6.

\bibitem{pedatzur2015attosecond}
O~Pedatzur, G~Orenstein, V~Serbinenko, H~Soifer, BD~Bruner, AJ~Uzan,
  DS~Brambila, AG~Harvey, L~Torlina, F~Morales, O~Smirnova, and N~Dudovich.
\newblock Attosecond tunnelling interferometry.
\newblock {\em Nature Physics}, 11(10):815, 2015.

\bibitem{eckle2008attosecond}
P~Eckle, AN~Pfeiffer, C~Cirelli, A~Staudte, R~Dorner, HG~Muller, M~Buttiker,
  and U~Keller.
\newblock Attosecond ionization and tunneling delay time measurements in
  helium.
\newblock {\em science}, 322(5907):1525--1529, 2008.

\bibitem{gianfrate2020measurement}
A~Gianfrate, O~Bleu, L~Dominici, V~Ardizzone, M~De~Giorgi, D~Ballarini,
  G~Lerario, KW~West, LN~Pfeiffer, DD~Solnyshkov, et~al.
\newblock Measurement of the quantum geometric tensor and of the anomalous hall
  drift.
\newblock {\em Nature}, 578(7795):381--385, 2020.

\bibitem{torma2023essay}
P{\"a}ivi T{\"o}rm{\"a}.
\newblock Essay: Where can quantum geometry lead us?
\newblock {\em Physical Review Letters}, 131(24):240001, 2023.

\bibitem{zak_berrys_1989}
J.~Zak.
\newblock Berry's phase for energy bands in solids.
\newblock 62(23):2747--2750.
\newblock Publisher: American Physical Society.

\bibitem{uzan-narovlansky_observation_2024}
Ayelet~J. Uzan-Narovlansky, Lior Faeyrman, Graham~G. Brown, Sergei Shames,
  Vladimir Narovlansky, Jiewen Xiao, Talya Arusi-Parpar, Omer Kneller, Barry~D.
  Bruner, Olga Smirnova, Rui E.~F. Silva, Binghai Yan, Álvaro Jiménez-Galán,
  Misha Ivanov, and Nirit Dudovich.
\newblock Observation of interband berry phase in laser-driven crystals.
\newblock 626(7997):66--71.
\newblock Publisher: Nature Publishing Group.

\bibitem{chacon_circular_2020}
Alexis Chacón, Dasol Kim, Wei Zhu, Shane~P. Kelly, Alexandre Dauphin, Emilio
  Pisanty, Andrew~S. Maxwell, Antonio Picón, Marcelo~F. Ciappina, Dong~Eon
  Kim, Christopher Ticknor, Avadh Saxena, and Maciej Lewenstein.
\newblock Circular dichroism in higher-order harmonic generation: Heralding
  topological phases and transitions in chern insulators.
\newblock 102(13):134115.
\newblock Publisher: American Physical Society.

\bibitem{yue2020imperfect}
Lun Yue and Mette~B Gaarde.
\newblock Imperfect recollisions in high-harmonic generation in solids.
\newblock {\em Physical Review Letters}, 124(15):153204, 2020.

\bibitem{vampa2015linking}
G~Vampa, TJ~Hammond, N~Thir{\'e}, BE~Schmidt, F~L{\'e}gar{\'e}, CR~McDonald,
  T~Brabec, and PB~Corkum.
\newblock Linking high harmonics from gases and solids.
\newblock {\em Nature}, 522(7557):462, 2015.

\bibitem{ghimire2011observation}
Shambhu Ghimire, Anthony~D DiChiara, Emily Sistrunk, Pierre Agostini, Louis~F
  DiMauro, and David~A Reis.
\newblock Observation of high-order harmonic generation in a bulk crystal.
\newblock {\em Nature physics}, 7(2):138, 2011.

\bibitem{luu2015extreme}
Tran~Trung Luu, M~Garg, S~Yu Kruchinin, Antoine Moulet, M~Th Hassan, and
  Eleftherios Goulielmakis.
\newblock Extreme ultraviolet high-harmonic spectroscopy of solids.
\newblock {\em Nature}, 521(7553):498, 2015.

\bibitem{schubert2014sub}
Olaf Schubert, Matthias Hohenleutner, Fabian Langer, Benedikt Urbanek, C~Lange,
  U~Huttner, D~Golde, T~Meier, M~Kira, Stephan~W Koch, et~al.
\newblock Sub-cycle control of terahertz high-harmonic generation by dynamical
  bloch oscillations.
\newblock {\em Nature Photonics}, 8(2):119, 2014.

\bibitem{schultze2014attosecond}
Martin Schultze, Krupa Ramasesha, CD~Pemmaraju, SA~Sato, D~Whitmore, A~Gandman,
  James~S Prell, LJ~Borja, D~Prendergast, K~Yabana, et~al.
\newblock Attosecond band-gap dynamics in silicon.
\newblock {\em Science}, 346(6215):1348--1352, 2014.

\bibitem{langer2018lightwave}
Fabian Langer, Christoph~P Schmid, Stefan Schlauderer, Martin Gmitra, Jaroslav
  Fabian, Philipp Nagler, Christian Sch{\"u}ller, Tobias Korn, PG~Hawkins,
  JT~Steiner, et~al.
\newblock Lightwave valleytronics in a monolayer of tungsten diselenide.
\newblock {\em Nature}, 557(7703):76--80, 2018.

\bibitem{tyulnev2024valleytronics}
Igor Tyulnev, {\'A}lvaro Jim{\'e}nez-Gal{\'a}n, Julita Poborska, Lenard Vamos,
  Philip St~J Russell, Francesco Tani, Olga Smirnova, Misha Ivanov, Rui~EF
  Silva, and Jens Biegert.
\newblock Valleytronics in bulk mos2 with a topologic optical field.
\newblock {\em Nature}, 628(8009):746--751, 2024.

\bibitem{mitra2024light}
Sambit Mitra, {\'A}lvaro Jim{\'e}nez-Gal{\'a}n, Mario Aulich, Marcel Neuhaus,
  Rui~EF Silva, Volodymyr Pervak, Matthias~F Kling, and Shubhadeep Biswas.
\newblock Light-wave-controlled haldane model in monolayer hexagonal boron
  nitride.
\newblock {\em Nature}, pages 1--6, 2024.

\bibitem{kneller_look_2022}
Omer Kneller, Doron Azoury, Yotam Federman, Michael Krüger, Ayelet~J. Uzan,
  Gal Orenstein, Barry~D. Bruner, Olga Smirnova, Serguei Patchkovskii, Misha
  Ivanov, and Nirit Dudovich.
\newblock A look under the tunnelling barrier via attosecond-gated
  interferometry.
\newblock 16(4):304--310.
\newblock Publisher: Nature Publishing Group.

\bibitem{garg2018ultimate}
Manish Garg, Hee-Yong Kim, and Eleftherios Goulielmakis.
\newblock Ultimate waveform reproducibility of extreme-ultraviolet pulses by
  high-harmonic generation in quartz.
\newblock {\em Nature Photonics}, 12(5):291--296, 2018.

\bibitem{hammond2017producing}
TJ~Hammond, DM~Villeneuve, and PB~Corkum.
\newblock Producing and controlling half-cycle near-infrared electric-field
  transients.
\newblock {\em Optica}, 4(7):826--830, 2017.

\bibitem{luu2018measurement}
Tran~Trung Luu and Hans~Jakob W{\"o}rner.
\newblock Measurement of the berry curvature of solids using high-harmonic
  spectroscopy.
\newblock {\em Nature communications}, 9(1):1--6, 2018.

\bibitem{luu2018observing}
Tran~Trung Luu and Hans~Jakob W{\"o}rner.
\newblock Observing broken inversion symmetry in solids using two-color
  high-order harmonic spectroscopy.
\newblock {\em Physical Review A}, 98(4):041802, 2018.

\bibitem{dudovich2006measuring}
N~Dudovich, Olga Smirnova, J~Levesque, Yu~Mairesse, M~Yu Ivanov, DM~Villeneuve,
  and Paul~B Corkum.
\newblock Measuring and controlling the birth of attosecond xuv pulses.
\newblock {\em Nature physics}, 2(11):781, 2006.

\bibitem{uzan2020attosecond}
Ayelet~Julie Uzan, Gal Orenstein, {\'A}lvaro Jim{\'e}nez-Gal{\'a}n, Chris
  McDonald, Rui~EF Silva, Barry~D Bruner, Nikolai~D Klimkin, Valerie Blanchet,
  Talya Arusi-Parpar, Michael Kr{\"u}ger, et~al.
\newblock Attosecond spectral singularities in solid-state high-harmonic
  generation.
\newblock {\em Nature Photonics}, 14(3):183--187, 2020.

\bibitem{qian_role_2022}
Chen Qian, Chao Yu, Shicheng Jiang, Tan Zhang, Jiacheng Gao, Shang Shi, Hanqi
  Pi, Hongming Weng, and Ruifeng Lu.
\newblock Role of shift vector in high-harmonic generation from
  noncentrosymmetric topological insulators under strong laser fields.
\newblock 12(2):021030.
\newblock Publisher: American Physical Society.

\bibitem{vampa_merge_2017}
G.~Vampa and T.~Brabec.
\newblock Merge of high harmonic generation from gases and solids and its
  implications for attosecond science.
\newblock 50(8):083001.
\newblock Publisher: {IOP} Publishing.

\bibitem{langer2017symmetry}
Fabian Langer, Matthias Hohenleutner, U~Huttner, Stephan~W Koch, M~Kira, and
  Rupert Huber.
\newblock Symmetry-controlled temporal structure of high-harmonic carrier
  fields from a bulk crystal.
\newblock {\em Nature photonics}, 11(4):227--231, 2017.

\bibitem{schmid2021tunable}
Christoph~P Schmid, Leonard Weigl, P~Gr{\"o}ssing, Vanessa Junk, Cosimo Gorini,
  Stefan Schlauderer, Suguru Ito, Manuel Meierhofer, Niklas Hofmann, Dmitry
  Afanasiev, et~al.
\newblock Tunable non-integer high-harmonic generation in a topological
  insulator.
\newblock {\em Nature}, 593(7859):385--390, 2021.

\bibitem{bai2021high}
Ya~Bai, Fucong Fei, Shuo Wang, Na~Li, Xiaolu Li, Fengqi Song, Ruxin Li, Zhizhan
  Xu, and Peng Liu.
\newblock High-harmonic generation from topological surface states.
\newblock {\em Nature Physics}, 17(3):311--315, 2021.

\end{thebibliography}
\bibliographystyle{unsrt}

\newpage
	\section*{Acknowledgments}
	We would like to acknowledge Shinsei Ryu and Tobias Holder for useful discussions and their scientific input. We also would like to thank Pengjie Wang for his great advice.
	N.D. is the incumbent of the Robin Chemers Neustein Professorial Chair. N.D. acknowledges the Minerva Foundation, the Israeli Science Foundation and the European Research Council for financial support. M.I. acknowledges funding of the DFG QUTIF grant IV152/6-2. A.J.G. acknowledges support from Comunidad de Madrid through TALENTO Grant 2022-T1/IND-24102. A.J.G. and M.I. acknowledge funding from the European Union’s Horizon 2020 research and innovation program under grant agreement No 899794. E. B. M. and R. E. F. S. acknowledge support from the fellowship LCF/BQ/PR21/11840008 from “La Caixa” Foundation (ID 100010434) and by Grant PID2021-122769NB-I00 funded by MCIN/AEI/10.13039/501100011033.

	\section*{Contributions}

	N.D. and A.J.U.N supervised the study. L.F. and A.J.U.N. conceived and planned the experiments. E.B.M, A.J.G. and R.S. performed the theoretical study and the numerical analysis. V.N., O.S. and M.I. developed the theoretical model. B.Y. provided the DFT calculations. L.F., A.J.U.N., T.A.P. and B.D.B performed the measurements. L.F., R.W. and R.P. analyzed the data. All authors discussed the results and contributed to writing the manuscript.
	\section*{Competing financial interests}
	The authors declare no competing financial interests.
	
	\section*{Corresponding authors}
	Correspondence to Nirit Dudovich (nirit.dudovich@weizmann.ac.il) and Ayelet Uzan-Narovlansky (auzan@princeton.edu).

        \section*{Data availability}
        The data and datasets that support the plots within this paper and other findings of this study are available from the corresponding author upon reasonable request.

        \section*{Code availability}
        The custom code used for the current study has been described in previous publications, and parts of it can be made available from the corresponding author on reasonable request.

        \section*{Supplementary information}
         Supplementary Information is available for this paper.

\newpage
\begin{figure}[hbt!]
	\centering
	  \includegraphics[trim= 10 170 10 20,clip,width=1\textwidth]{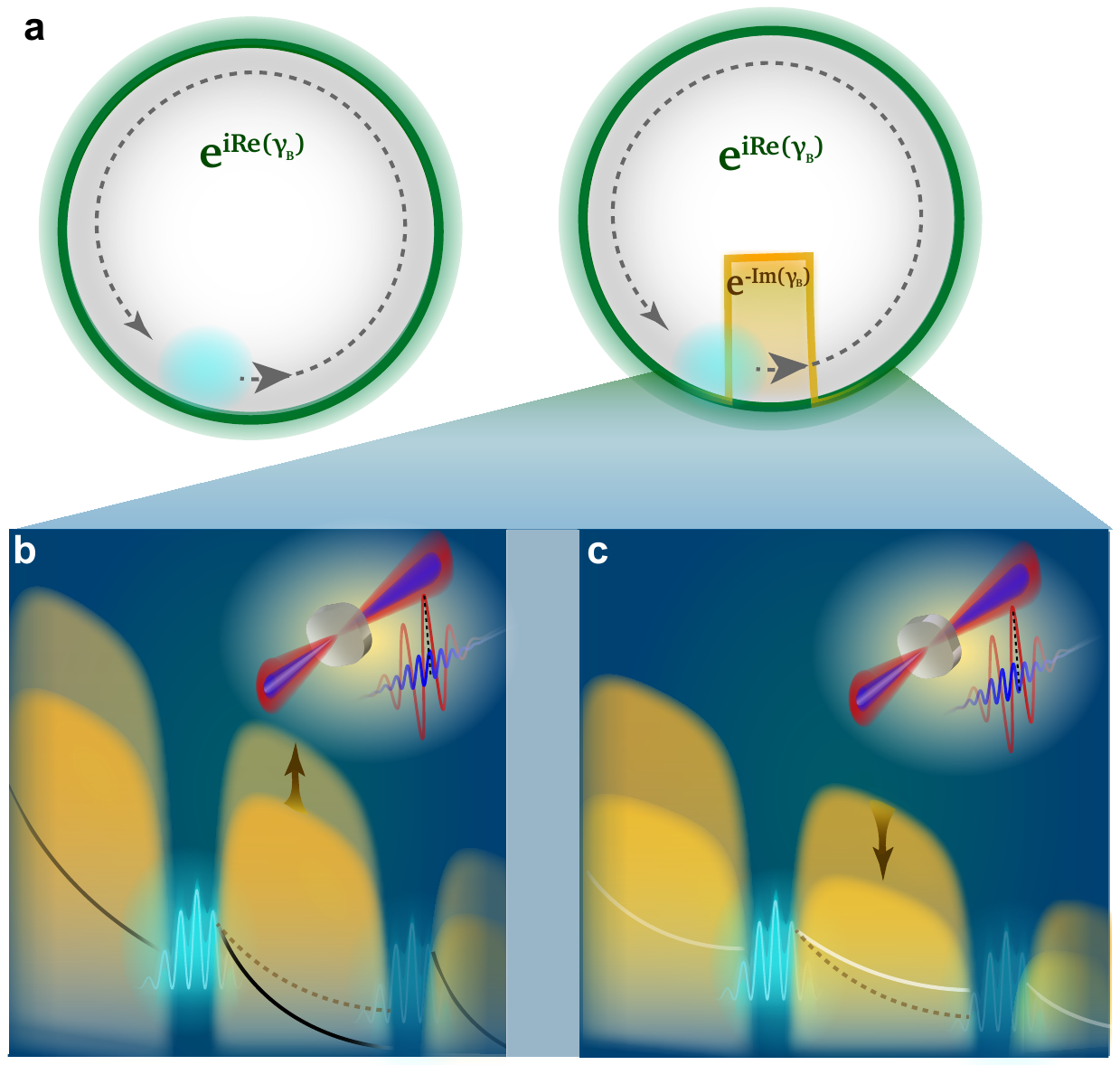}
\medskip
\caption{\textbf{Imaginary Berry phase under tunneling barrier. a,} When a quantum wavepacket follows a closed loop trajectory in the parameter space, it accumulates a real valued Berry phase (left). The presence of a tunneling barrier (right), introduces losses leading to the accumulation of a complex geometrical phase. \textbf{b and c,} Manipulation of the tunneling barrier in light driven crystals using a two color field. The strong laser field (red) bends the crystal Coulomb potential (orange), allowing the electronic wavefunction (cyan) to tunnel from the valence band to the conduction band (black/white lines). Adding the second harmonic field (dark blue) modulates the tunneling barrier, increasing or decreasing the barrier, as we scan the two-color delay (b or c respectively). }
	\label{fig1:fig1}
\end{figure}

\newpage
\begin{figure}[hbt!]
	\centering
	  \includegraphics[trim= 20 250 20 100,clip,width=1\textwidth]{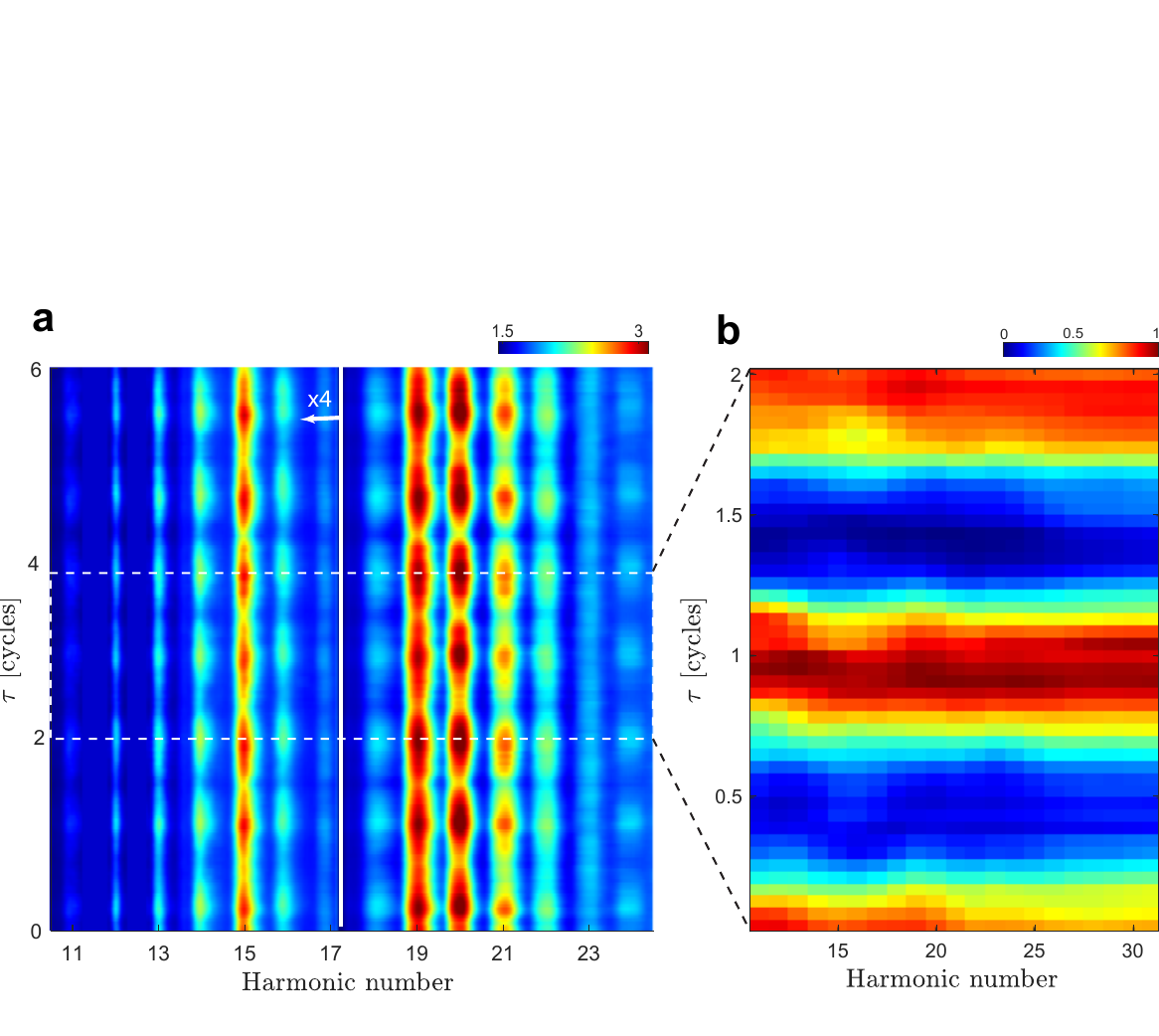}
\medskip
\caption{\textbf{Two color HHG spectroscopy in quartz. a,} Oscillating HHG spectrum (in log scale) in quartz, as a function of the two color delay (vertical axis), along $\Gamma-K$. \textbf{b} Averaged sum of adjacent even and odd harmonics, as a function of the two color delay (vertical axis), generated in quartz. The sum of adjacent harmonics is proportional to the imaginary perturbation induced by the SH field ($\sigma^i(\tau)$). Each harmonic signal is normalized by its maximum value and subtracted by the DC signal.
}
	\label{fig2:fig2}
\end{figure}

\newpage
\begin{figure}[hbt!]
	\centering
	  \includegraphics[trim= 50 215 0 50,clip,width=1\textwidth]{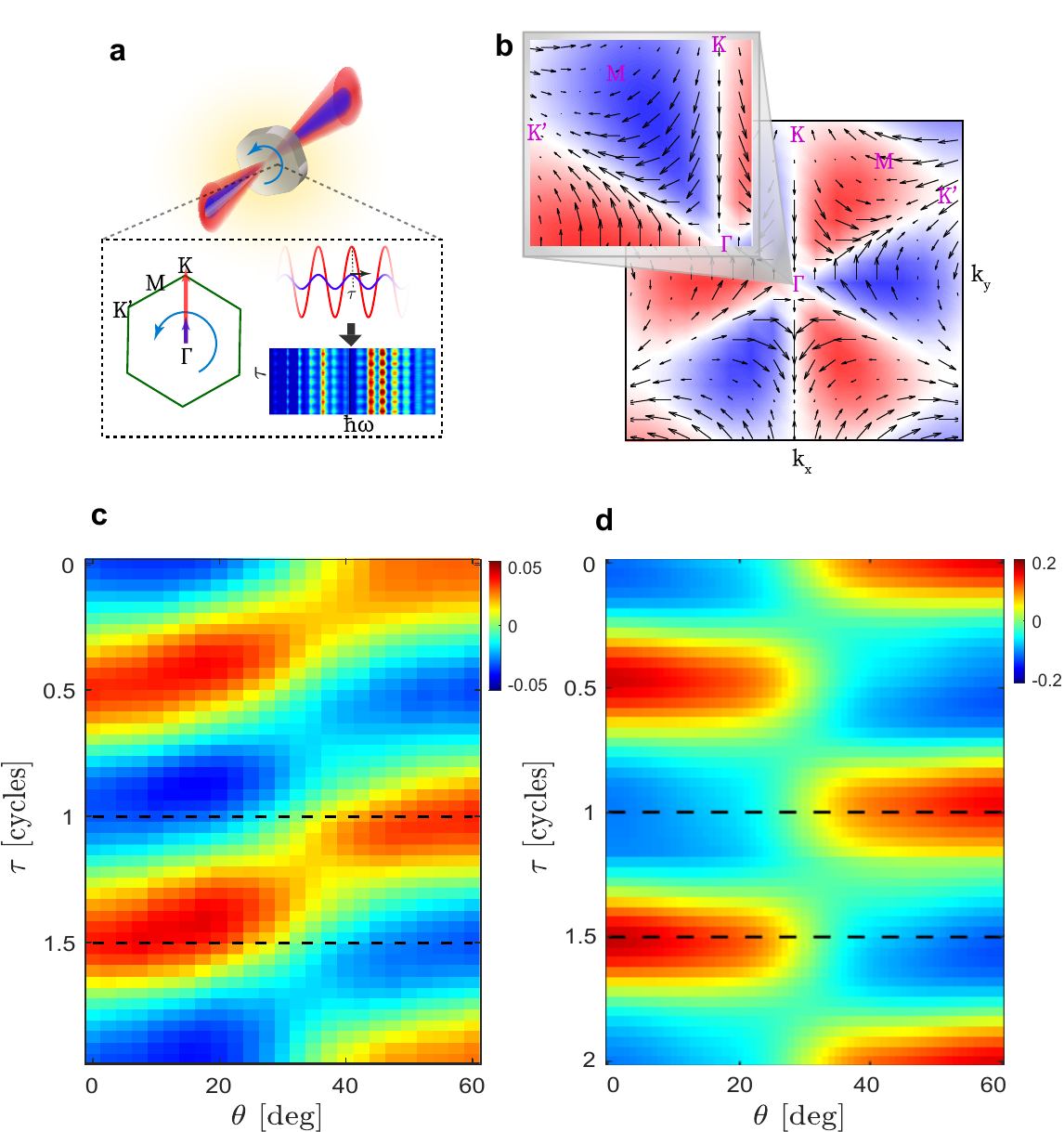}
\medskip
\caption{\textbf{Resolving the imaginary phase and its angular dependence. a,} By rotating the crystal with respect to the fundamental field's polarization, while scanning the SH delay, we resolve the angular dependence of the HHG oscillations ($0^\circ$ orientation represents $\Gamma-K$ and $60^\circ$ represents $\Gamma-K'$). \textbf{b,} Theoretical description (illustration) for the Berry connection (black arrows) in quartz, plotted on top of the Berry curvature (blue/red colors for positive/negative values). The presented Berry connection follows the crystal symmetries showing a sign changed between $\Gamma-K$ to $\Gamma-K'$. \textbf{c,d} Experimentally/theoretically resolved imaginary phase (H14/H11), $I_{sum}$, as a function of the SH delay and crystal orientation. This term probes the overall imaginary phase, revealing the interplay between the imaginary Berry phase ($tanh(\gamma^i_B)$) and the imaginary perturbation induced by the two color field ($\sigma^i$). Positive (negative) values signify suppression (amplification) of the imaginary phase, due to the corresponding modification of the tunneling barrier. Dashed lines (black) emphasize the $\pi$ phase shift between $\Gamma-K$ and $\Gamma-K'$ crystal orientations.}

	\label{fig3:fig3}
\end{figure}

\newpage
\begin{figure}[hbt!]
	\centering
	  \includegraphics[trim= 20 180 20 20,clip,width=1\textwidth]{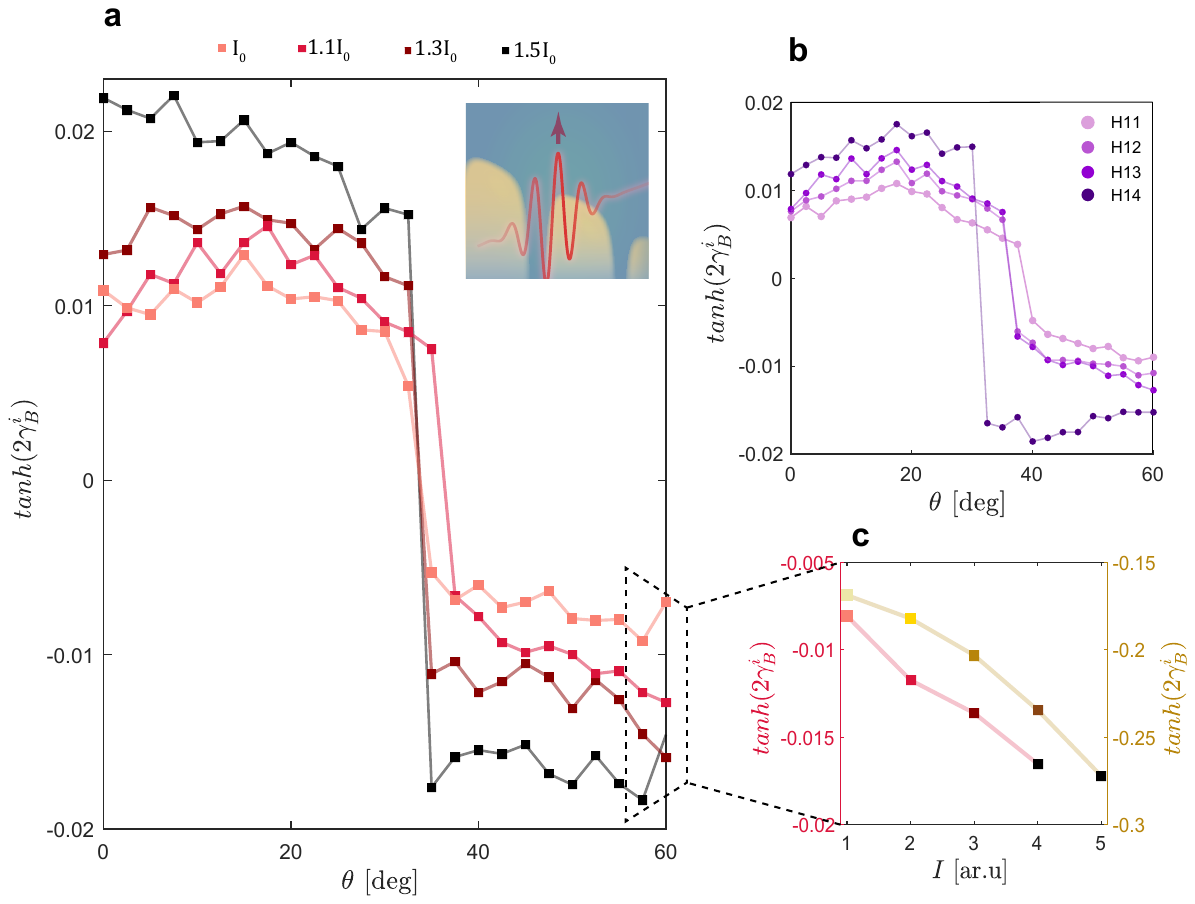}
\medskip
\caption{\textbf{Extracting the angular dependence of the imaginary Berry phase. a,} Field's intensity dependence. The imaginary Berry phase (H13), $\propto\tanh{(2\gamma^i_B)}$, resolved with different fundamental field intensities ($I_0$-$1.5I_0$, where $I_0$ is the lowest intensity and it is of the order of $10^{13}\frac{W}{cm^2}$) as a function of the crystal orientation. \textbf{b,} Extracting the evolution of the imaginary Berry phase within the optical cycle. The imaginary Berry phase, $\propto\tanh{(2\gamma^i_B)}$, extracted for different harmonic orders, as a function of the crystal orientation. \textbf{c,} Measured/calculated imaginary Berry phase along ($\Gamma-K'$), for different field intensities (red/yellow respectively), For the experimental plots, 1-4 intensities (x axis) are as in figure \textbf{a} and for the theoretical calculation 1-5 intensities are 1-1.5 $\frac{TW}{cm^2}$.}
 
	\label{fig3:fig3}
\end{figure}

\end{document}